\documentstyle[12pt,epsf]{article}
\begin{document}

\newcommand{\sheptitle}
{The Fate of Dark Energy}
\newcommand{\shepauthor}
{Paul H. Frampton and Tomo Takahashi}
\newcommand{\shepaddress}
{Department of Physics and Astronomy,\\
University of North Carolina, Chapel Hill, NC 27599-3255.}
\newcommand{\shepabstract}
{}

\begin{titlepage}
\begin{flushright}
astro-ph/0211544\\
\today
\end{flushright}
\vspace{.1in}
\begin{center}
{\large{\bf \sheptitle}}
\bigskip \medskip \\ \shepauthor \\ \mbox{} \\ {\it \shepaddress} \\
\vspace{.5in}

\bigskip \end{center} \setcounter{page}{0}
\shepabstract
\begin{abstract}
By studying the present cosmological data, particularly
on CMB, SNeIA and LSS, we find that the future fate of
the universe, for simple linear models of the dark energy
equation-of-state, can vary between the extremes
of (I) a divergence of the scale factor in as little as 7 Gyr; (II)
an infinite lifetime of the universe
with dark energy dominant for all future time;
(III) a disappearing dark energy where the
universe asymptotes as $t \rightarrow \infty$
to $a(t) \sim t^{2/3}$ {\it i.e.} matter domination.
Our dreadful conclusion is that no amount of data
from our past light-cone can select between these future scenarios.
\end{abstract}

\vspace*{6cm}
\begin{flushleft}
\hspace*{0.9cm}

\begin{tabular}{l} \\
\hline
{\small Emails: frampton@physics.unc.edu,
takahash@physics.unc.edu}

\end{tabular}
\end{flushleft}

\end{titlepage}

\newpage

\noindent {\it Introduction}

The cosmic concordance of data from three disparate sources:
Cosmic Microwave Background (CMB), Large Scale Structure (LSS)
and High-Red-Shift Supernovae (SNeIA) suggests that the
present values of the dark energy and matter components,
in terms of the critical density, are approximately
$\Omega_{\rm X} \simeq 0.7$ and $\Omega_{\rm M} \simeq 0.3$.
The question to which we try to make a small contribution
in this paper is to what extent precision
cosmological data will allow
us to discriminate between possible futute fates
of the Universe?

If one makes the most conservative assumption that
$\Omega_X$ corresponds to a cosmological constant
with Equation of State given by a constant
$w = p/\rho = -1$, then the future evolution of
the universe follows from the Friedmann
equation

\begin{equation}
\left( \frac{\dot{a}}{a} \right)^2 = \frac{8 \pi G \rho}{3} +
\frac{\Lambda}{3}
\label{friedmann}
\end{equation}
in which $a(t)$ is the scale factor normalized at the present time as
$a(t_0)=1$, $\rho(t)=\rho(0)a^{-3}$ is the energy density of matter
component and $\Lambda$ is constant.  Here we assumed that the universe
is flat and neglected radiation.

In such a simple, and still viable, case the behavior
of $a(t)$ for asymptotically large $t \rightarrow \infty$
is
\begin{equation}
a(t) \sim \exp \left( \sqrt {\frac{\Lambda}{3}} t \right)
\label{desitter}
\end{equation}
so that the dark energy asymptotically dominates
and the universe is blown apart in an infinite time
$a(t) \rightarrow \infty$ as $t \rightarrow \infty$.

Even assuming that $w$ is constant, however, there is a wide range
of possible $w$: according to \cite{HM} the allowed values are $-2.68
< w < -0.78$. We do not assume this result but will arrive at
a similar allowed range; the difference
is because our priors are slightly different 
(we fix the cosmological parameters 
$\Omega_M = 0.3, \Omega_{\Lambda} = 0.7,
\Omega_b = 0.02,$ and $h = 0.65$
instead of allowing them to vary.)
 The asymptotic behavior in Eq.(\ref{desitter}) is
very far from established by present data. 
The case $w < -1$ has the property
that boosting from the dark energy rest
frame to an inertial frame with velocity satisfying $(v/c)^2 > -1/w$
leads to a negative energy density,
but this does not
violate any law of physics\footnote{
This violates the weak energy condition\cite{hawk,Ford}.
}. 
Here we shall consider 
some simple models for $w$, including dependence on red-shift $w(z)$,
to illustrate how far existing data are from answering the
question of the future fate of the universe.  As we will show, for a
model in which $w$ varies linearly with red-shift, present data
are consistent with extremely different futures. For examples,
in one case the
scale factor diverges\cite{caldwell,starobinsky} in finite time
\footnote{ 
Gravitationally-bound
systems could survive longer than
$t_r$ in Eq.(\ref{tr}) but such systems would
be infinitely separated from one another.
},
in just another $7$ Gyr, while in another case the energy density of dark
energy decreases eventually faster than that of matter, i.e., the dark energy
disappears and the universe reverts to being matter-dominated with
$a(t) \sim t^{2/3}$, as $t \rightarrow \infty$.

\bigskip
\bigskip

\noindent {\it Constant Equation of State.}

\bigskip
\bigskip
Here we discuss the future fate of the universe in the case of the
constant equation of state.  If we assume, to begin, that $w$ is
constant then, keeping only the dark energy term
\begin{equation}
\left( \frac{\dot{a}}{a} \right)^2 = H_0^2 \Omega_X a^{-\beta}
\label{constantw}
\end{equation}
where $\beta = 3(1+w)$.  Most authors have discussed the case with $w
\ge -1$, however the case with $w < -1$ is also possible and
discussed phenomenologically in \cite{HM,caldwell,schulz},
and in connection with string theory in
\cite{PHF}.  If $\beta < 0$,
corresponding to $w < -1$, the solution of Eq.(\ref{constantw})
diverges at a finite time $t = t^{*}$. By integrating
\begin{equation}
\int^{\infty}_{a(t_0)} a^{\beta/2 -1} = H_0 \sqrt{\Omega_X}
\int^{t^{*}}_{t_0} dt
\label{integral}
\end{equation}
one finds that the remaining time $t_{\rm r}$ before time ends
$t_{\rm r} = (t^{*} - t_0)$ is given analytically by
\begin{equation}
t_{\rm r} = \frac{2}{3 H_0} \frac{1}{\sqrt{\Omega_X} (-w-1)}
\label{tr}
\end{equation}

In Eq.(\ref{tr}), putting in $\Omega_X = 0.7$ and
$\frac{2}{3} H_0^{-1} = 9.2$ Gyr one finds for
$w = -1.5, -2.0$ and $-2.5$, respectively
$t_{\rm r} =  22,$ 11 and 7.3 Gyr.

In such a constant $w$ scenario which is consistent
with all cosmological data, the divergence of the scale factor
will occur
in a finite time period of $7$ Gyr (or more)
from now.

With respect to the Solar System, this end of time occurs
generally after the Sun has transformed into a Red Giant,
and swallowed the Earth, as is
expected approximately $5$ Gyr in the future.

\bigskip
\bigskip

\noindent {\it Equation of State Varying Linearly with Red-Shift.}

\bigskip
\bigskip

As a more general ansatz, we consider the model for $w$ depending
linearly on red-shift:\footnote{
A model with $w$ linearly depending on red-shift is also
discussed in \cite{DW} but fitting the CMB data was not
investigated. Another parametrization of $w(Z)$
is in \cite{linder}.
}

\begin{equation}
w(Z) = w(0) + C Z \theta(\zeta - Z) + C \zeta \theta(Z -\zeta)
\label{EoS}
\end{equation}

\bigskip

\noindent where the modification is cut off arbitrarily at some $Z=\zeta >
0$.
We assume $C \leq 0$ and consider the two-dimensional
parameter space spanned by the two variables $w(0)$ and $C$.

In order to discuss constraints on our phenomenological model, we
compare its prediction with experimental data from SNeIA and CMB.
We evaluate the goodness-of-fit parameter $\chi^2$ as a function of
$w(0)$ and $C$.  For SNe1A, we used a dataset consisting of 37 SNe
from \cite{SN_Riess} with the MLCS method.  To calculate the CMB power
spectrum, we used a modified version of CMBFAST \cite{CMBFAST}. For
the analysis of CMB, we used experimental data from COBE \cite{COBE},
BOOMERanG \cite{BOOMERANG}, MAXIMA \cite{MAXIMA} and DASI \cite{DASI}.
To calculate $\chi^2$, we adopt the offset log-normal approximation
\cite{BJN} and used RADPACK package \cite{RADPACK}.  We also studied
the constraint from LSS using the 2dF data \cite{LSS}, and found that
it does not give severe constraint on the parameters $w(0)$ and $C$.

For an illustration, we take the cosmological parameters as
$\Omega_{\rm b}h^2=0.02, \Omega_{\rm M}=0.3, \Omega_{\rm X}=0.7$ and
$h=0.65$, and the initial power spectra are assumed to be scale
invariant in all numerical calculation in this paper.

To set the stage, let us first use only the SNe1A data to constrain
the parameters $w(0)$ and $C$. The result is shown for $\zeta = 2$ in
Figure 1 where the 99 \% C. L. allowed region is the region between
the two dashed lines shown. We may remark three distinct regions:

\noindent (I) $w(0) < (C - 1)$. In this case there is divergence of
the scale factor, at a finite future time.

\noindent (II) $(C - 1) \leq w(0) < C$. Here the lifetime of the universe
is infinite. The dark energy dominates
over matter, as now, at all future times.

\noindent (III) $C \leq w(0)$. The lifetime of the universe is again
infinite but after a finite time the dark energy will disappear
relative to the dark matter and matter-domination will be
re-established with $a(t) \sim t^{2/3}$.

\bigskip
\bigskip

When we add the constraints imposed by the CMB data, the allowed
region is smaller as shown in Figure 2, plotted for $\zeta =
0.5$. Such a small $\zeta$ still allows all three future possibilities
(I), (II) and (III).  For somewhat larger $\zeta$ only possibilities
(I) and (II) are allowed in this particular parameterization.

\bigskip
\bigskip

The case $\zeta = 2$ is exhibited in more detail for different values
of $w(0)$ and $C$ in Figures 3 and 4.  Figure 3 shows the variation of
the transition red-shift $Z_{\rm tr}$ where deceleration changes to
accelerated cosmic expansion defined by $q(Z_{\rm tr}) = 0$. From the
figure, we can read off that $Z_{\rm tr}$ becomes smaller as $C$
becomes more negative for fixed $w(0)$; this is because the epoch where dark
energy becomes the dominant component of the Universe becomes later.  
This affects the magnitude-red shift relation of high-Z
supernovae.  In Figure 4, the magnitude-red shift relation is shown
for the SNeIA data \cite{SN_Riess} along with the prediction of our
phenomenological model for $\zeta = 2$.  The magnitudes are calculated,
as usual, relative to the empty universe Milne model with
$\Omega_M =0, ~~ \Omega_X = 0$ and $\Omega_k = 1$.  One can expect that
high-redshift SNe would appear dimmer if $C$ were more negative.

\bigskip
\bigskip

To return to our main point, let us assume that more precise cosmological
data
will allow an approximate determination of $w(Z) = f(Z)$ as a function of
$Z$
for positive $Z > 0$. Then to illustrate the possible future evolutions
write:

\begin{equation}
w(Z) = f(Z)\theta(Z) + (f(0) + \alpha Z) \theta(-Z)
\label{EoS2}
\end{equation}
In this case, the future scenarios
(I), (II) and (III) occur respectively
for $\alpha > (f(0)+ 1)$, $(f(0) + 1) > \alpha > f(0)$
and $\alpha < f(0)$.

Present data are consistent with a simple cosmological constant
$f(Z) = -1$ in Eq.(\ref{EoS2}) in which
case the divergence of the scale factor 
occurs for $\alpha > 0$, the infinite-time
dark energy domination for $0 > \alpha > -1$, and
disappearing dark energy for $\alpha < -1$.

\bigskip

Since in practice $F(Z)$ for $Z \geq 0$ will never be determined
with perfect accuracy the continuation of $w(Z)$
to future $Z < 0$ will be undecidable from observation as will therefore
be the ultimate fate of the Universe.

\bigskip
\bigskip
\bigskip

\noindent {\it Acknowledgment.}

This work was supported in part by the US Department of Energy
under Grant No. DE-FG02-97ER-41036.
We acknowledge the use of CMBFAST \cite{CMBFAST} and RADPACK
\cite{RADPACK}.

\bigskip

\newpage

\noindent {\bf Figure Captions}

\bigskip

\noindent Figure 1. Supernovae 1A constraints for $\zeta = 2$ showing
the three regions (I), (II) and (III) with different futures as
discussed in the text.

\bigskip

\noindent Figure 2. Addition of the CMB data can permit all three
regions for sufficiently small $\zeta$, only (I) and (II) for larger
$\zeta$. This plot is for small $\zeta = 0.5$.

\bigskip

\noindent Figure 3. $Z_{\rm trans}$, defined in the text, is plotted
as a function of $w(0)$ for any $\zeta \geq 0.7$.

\bigskip

\noindent Figure 4. The SNeIA data and predicted curve of our
phenomenological models for $\zeta = 2$  are plotted relative
to an empty universe Milne model with $\Omega_M = \Omega_X = 0$
and $\Omega_k = 1$.

\newpage

\bigskip

\begin{figure}

\begin{center}

\epsfxsize=7.0in
\ \epsfbox{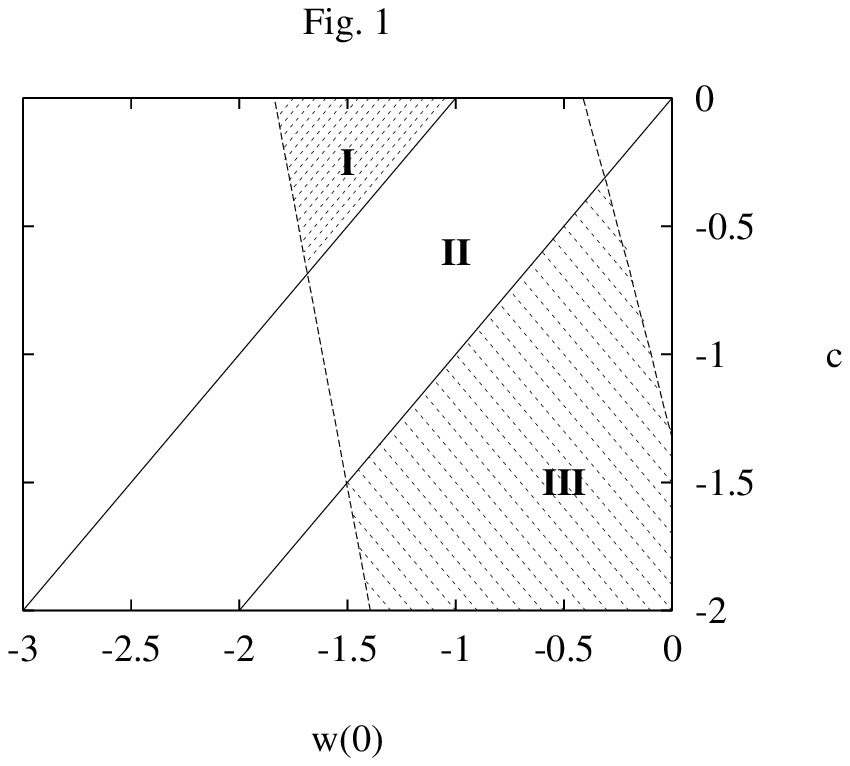}

\end{center}

\end{figure}

\newpage

\begin{figure}

\begin{center}

\epsfxsize=7.0in
\ \epsfbox{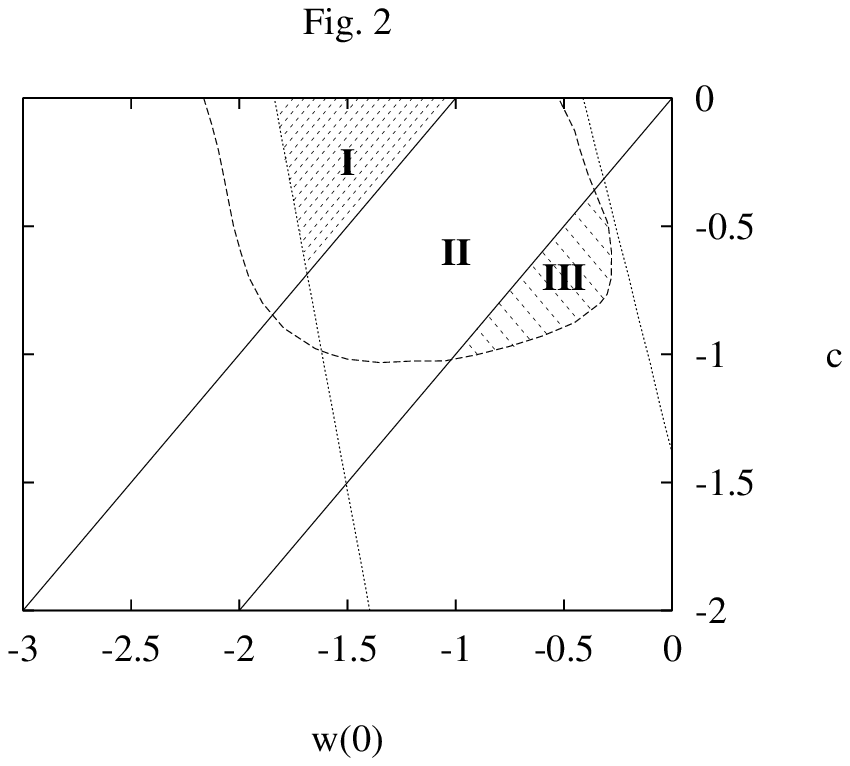}

\end{center}
\end{figure}

\newpage

\begin{figure}

\begin{center}

\epsfxsize=5.0in
\ \epsfbox{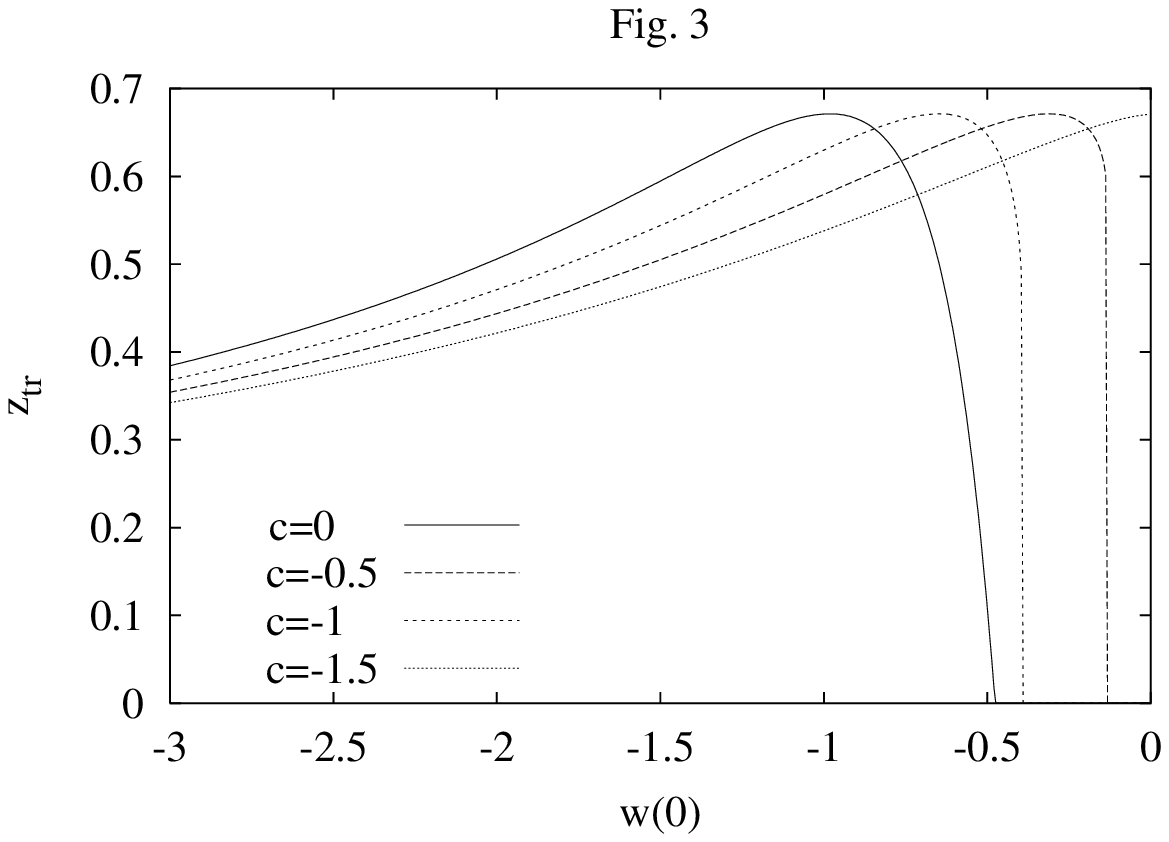}

\end{center}
\end{figure}

\newpage

\begin{figure}

\begin{center}

\epsfxsize=5.0in
\ \epsfbox{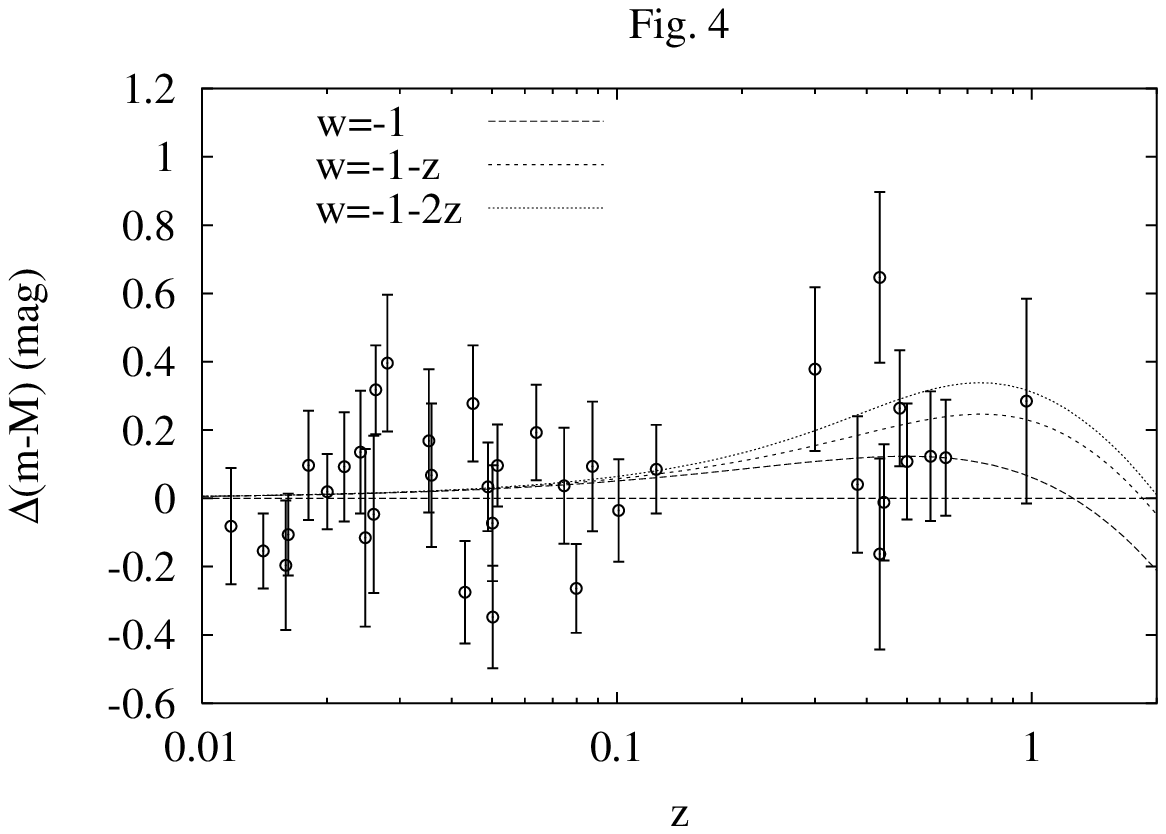}

\end{center}
\end{figure}

\end{document}